\documentclass{osa-article}
    \usepackage{braket}

\journal{oe}
\articletype{Research Article}
\begin{document}

\title{Open-Loop Polarization Mode Dispersion Mitigation for Fibre-Optic Time and Frequency Transfer}

%\author{Thomas Fordell\authormark{1,*}}% Author Two,\authormark{2,*} and Author Three\authormark{2,3}}
\author{Thomas Fordell\authormark{*}}% Author Two,\authormark{2,*} and Author Three\authormark{2,3}}
\address{VTT Technical Research Centre of Finland Ltd, National Metrology Institute VTT MIKES, Tekniikantie 1, Espoo, P.O.Box 1000, FI-02044 VTT, Finland\\}
%\address{\authormark{1}VTT Technical Research Centre of Finland Ltd, National Metrology Institute VTT MIKES, Tekniikantie 1, Espoo, FI-02044 VTT, Finland\\}
%\authormark{2}Publications Department, The Optical Society (OSA), 2010 Massachusetts Avenue NW, Washington, DC 20036, USA\\
%\authormark{3}Currently with the Department of Electronic Journals, The Optical Society (OSA), 2010 Massachusetts Avenue NW, Washington, DC 20036, USA}
\email{$^*$thomas.fordell@vtt.fi} %% email address is required



\begin{abstract}
The non-reciprocal and dynamic nature of polarization mode dispersion (PMD) in optical fibers can be a problem for accurate time and frequency transfer. Here a simple, passive solution is put forward that is based on transmitting  optical pulses with alternating orthogonal polarization. The fast and deterministic polarization modulation means that the PMD noise is pushed far away from the frequencies of interest and upon reflection from a Faraday mirror at the receiver, the pulses have a well defined polarization when they return to the transmitter, which facilitates stable optical phase detection and fibre phase compensation. In an open-loop test setup that uses a mode-locked laser and a simple pulse interleaver, the polarization mode dispersion is shown to be reduced by more than two orders of magnitude.\\ 
\\
© 2022 Optica Publishing Group. Users may use, reuse, and build upon the article, or use the article for text or data mining, so long as such uses are for non-commercial purposes and appropriate attribution is maintained. All other rights are reserved.\\
\href{https://doi.org/10.1364/OE.448553}{https://doi.org/10.1364/OE.448553}
\end{abstract}

%%%%%%%%%%%%%%%%%%%%%%%%%%  body  %%%%%%%%%%%%%%%%%%%%%%%%%%
\section{Introduction}

%The fractional frequency instability of optical oscillators has reached the $10^{-17}$ regime \cite{Hafner2015a,Matei2017a}, optical clockwork can translate this performance to a very wide frequencies range \cite{Benkler2019a,Giunta2020a}, and the fractional uncertainty of optical frequency standards is currently at the $10^{-18}$ level \cite{Brewer2019a, Huntemann2016a,McGrew2018a, Bothwell2019a}. 

The impressive development of optical oscillators  \cite{Hafner2015a,Matei2017a}, optical clockwork \cite{Benkler2019a,Giunta2020a} and optical frequency standards \cite{Brewer2019a, Huntemann2016a,McGrew2018a, Bothwell2019a} puts high-demands on signal distribution, since frequency comparisons can be used besides time keeping for interesting applications in, e.g.,  relativistic geodesy and the search for new physics. %\cite{Takano2016a,Grotti2018a,Mehlstaubler2018a,Delva2019a,Huang2020a} 
%and the search for new physics \cite{Blatt2008a,Godun2014a, Huntemann2014a,Kennedy2020a,Roberts2020a, Lange2021a, Beloy2021a}. 
Mode-locked lasers have been shown to possess extremely low phase noise \cite{Song2011a,Benedick2012a,Kalubovilage2020a}, enabling tight time synchronization in the attosecond regime \cite{Kim2007a,Xin2014a}. With high-power, high-speed photodetectors the performance can be carried over to the microwave regime \cite{Baynes2015a,Xie2017a,Hyun2020a,Nakamura2020a} where very-long baseline interferometry, future telecommunications systems and time-resolved ultrafast x-ray studies can benefit from low-phase-noise sources and precise synchronization. Because of this a considerable amount of effort has been put into fiber-optic time  \cite{Kim2007a,Marra2012a,WRproject,Sliwczynski2013a,Xin2014a,Jung2014a,Kodet2015a,Dierikx2016a,Nagano2016a,Lessing2017a,Rizzi2018a,Safak2018a,Abuduweili2020a}, optical frequency \cite{Predehl2012,Bai2013a,Bercy2014a,Grosche2014a,Lisdat2016a,Guillou-Camargo2018a,Clivati2015a}, and RF frequency \cite{Lopez2008a,Williams2008a,Lopez2010a,Hsu2012a,Gao2012s,He2013a,Wang2014a,Zhang2014a,Wang2015a,Zhang2015a,Schediwy2017a,He2018a} distribution technologies and combinations thereof.   

% https://www.osapublishing.org/oe/fulltext.cfm?uri=oe-29-10-14505&id=450550

Accurate time and frequency transfer relies on signals travelling in both directions so that changes in the transmission medium can be detected and compensated for in real time at the transmitter or in post processing. For this to work, the signals should experience the same delay in both directions. Unfortunately, in single-mode fibers used for long-haul communications, polarization is not preserved and different polarizations experience different delays due to fiber imperfections. Small, random birefringence in the fibers lead to polarization mode dispersion (PMD), which for modern fibers is $\lesssim 0.1$~ps$/\sqrt{\textrm{km}}$. PMD is non-reciprocal and dynamic and difficult to compensate for using, e.g., polarization controllers. PMD compensation techniques have been investigated considerable up until the turn of the century \cite{Galtarossa2005} since PMD was expected to hinder data transmission at several tens of Gbps. The move to coherent communication systems has lessened emphasize on PMD compensation since its effect can, in principle, be completely compensated for with digital signal processing in a coherent receiver \cite{Xie2011}. 
In \cite{Clivati2020a}, a dual-polarization coherent receiver was shown to be effective against beat note signal fading due to polarization changes in optical frequency dissemination.

%h fading of optical beat notes due to polarization changes, a dual-polarization coherent receiver was recently tested for optical frequency dissemination in \cite{Clivati2020a}.

Nevertheless, for precise time and frequency distribution, the non-reciprocal and time dependant nature of PMD remains an issue. It has been observed in, e.g., femtosecond optical time transfer over 300~m\cite{Sydlo2020} and in RF transfer over a 2-km link \cite{Jung2014a}, over a 2x50-km link \cite{Fordell2020a} and over an 86-km urban link \cite{Lopez2008a, Lopez2010a}. In \cite{Zhang2013a}, %bidirectional time division multiplexing transmission over a single fiber with the same wavelength was analyzed and 
it was concluded that PMD may become a considerable obstacle for ultra-long-haul fiber-optic time transfer. Very recently non-reciprocity was analyzed in \cite{Xu2021a} and it was determined that PMD is the dominant noise mechanism at low frequencies. In \cite{Gibbon2015a}, PMD was studied for radio telescope synchronization, and it was concluded that while it is not an issue for the MeerKAT telescope, it remains a challenge for the planned Square Kilometer Array. It should be noted that PMD is not always an issue \cite{Williams2008a}; its magnitude and relevance depend on the specifications of the used fibres, the dynamical properties (stability) of the environment, the time span of the experiment and, of course, the performance requirements of the link. 

Two approaches have been used to overcome polarization mode dispersion: polarization maintaining (PM) fibers and polarization scrambling. Since PM fibers have higher losses and are expensive, they have only been used for fairly short distances \cite{Xin2014a, Safak2018a}. Polarization scrambling is another option, which can be done in several ways. Extremely fast scrambling on a great circle of the Poincar\'e sphere can be done with a single electro-optic phase modulator \cite{Kersey1990a, Zhang2014a}. Another option is to, e.g., use several piezo-electric fiber squeezers in series to get full coverage of the Poincar\'e sphere\cite{Yao2002a}.
 
The present work investigates the use of light pulses from a mode-locked laser with alternating orthogonal polarizations for mitigating PMD. In the next Section it is shown that the average delay experienced by two pulses with orthogonal polarizations does not suffer from PMD, and in Section 3 this is demonstrated experimentally in a test setup where the PMD is reduced by more than two orders of magnitude. For high-speed fiber-optic telecommunications, alternating between two orthogonal polarization has been considered for a long time, e.g., for increasing spectral efficiency and for reducing intra-channel and inter-channel non-linearities due to four-wave mixing and cross-phase modulation \cite{Evangelides1992,Matera1999a, Ito1999a, Hodzic2003a, Xie2004a}. Many other polarization modulation and scrambling schemes have been considered \cite{Bergano1996a, Park2008}, but to the best of the author's knowledge, the use of time-interleaved orthogonal polarisation states for PMD mitigation in time and frequency distribution has not been investigated before.

\section{Theory}

%Following \cite{Shieh1999a} and \cite{Poole1986a}, 
Following the treatment of PMD in \cite{Shieh1999a}, which extends the results of \cite{Poole1986a}, let the complex electric field of a light pulse be described by the Jones vector  $\ket{E} = \binom{E_x(t)}{E_y(t)}$. The temporal position of the pulse is then given by
\begin{equation}
\langle t\rangle = \frac{\int E_x^* t E_x dt + E_y^* t E_y dt}{\int |E_x|^2 + |E_y|^2 dt}= \frac{1}{A}\int \bra{E}t\ket{E} dt = \frac{i}{A}\int \bra{\tilde{E}}\partial_\omega\ket{\tilde{E}}d\omega,
\end{equation}    
where $A=\int \langle E \vert E \rangle dt $ and  $\ket{\tilde{E}}$ is the Fourier transform of $\ket{E}$. If the medium (optical fiber) is taken to be linear, the input field $\ket{\tilde{E}_1}$ is transformed upon propagation to 
\begin{equation}
\ket{\tilde{E}_2} = T\ket{\tilde{E}_1} = BU\ket{\tilde{E}_1}, 
\end{equation}
where the transfer matrix $T=T(\omega)$ has been split into a  scalar $B=B(\omega)$ containing polarization independent losses and dispersion and a matrix $U=U(\omega)$ containing polarization dependent effects. If %$U$ acts as an arbitrary phase retarder 
 there is no polarization dependent gain or loss, then $U$ is unitary ($U^\dagger U = 1$). The temporal position of $\ket{\tilde{E}_2}$ is then given by

\begin{eqnarray}
\langle t\rangle &=& \frac{i}{A}\int \bra{\tilde{E}_1}U^\dagger B^*\partial_\omega B U\ket{\tilde{E}_1}d\omega \nonumber \\ &=&  t_0 + \frac{i}{A} \int \vert B\vert^2 \bra{\tilde{E}_1}U^\dagger (\partial_\omega U) \ket{\tilde{E}_1} d\omega,
\end{eqnarray}
where $t_0$ contains the polarization independent part. This corresponds to Eqs. 6-8 in\cite{Shieh1999a}, which proceeds to discuss the (generalized) principal states of polarization as introduced in \cite{Poole1986a}. It is shown that there are orthogonal input states of polarization that minimize and maximize the delay of a pulse. At the output of the fiber, these two states are also orthogonal to each other.

If we, instead of the principal states of polarization, simply consider the average arrival time of two similar but orthogonal, non-overlapping pulses $\ket{E_{11}}$ and $\ket{E_{12}}$ such that

\begin{equation}
\ket{E_{11}} = E(t)\ket{\rho_1}, \ket{E_{12}} = E(t-\tau)\ket{\rho_2} \textrm{and } \langle \rho_i \vert \rho_j \rangle =\delta_{ij},
\end{equation}
and if the transmitter has a constant (or very slowly varying) polarization, then $\ket{\rho_i}$ can be considered constant and $\ket{\tilde{E}_{11}} =\tilde{E}(\omega) \ket{\rho_1}$, $\ket{\tilde{E}_{12}} =e^{-i\omega \tau}\tilde{E}(\omega) \ket{\rho_2}$ yielding

\begin{equation}
    \langle t \rangle = \frac{1}{2}(\langle t_1 \rangle + \langle t_2  \rangle) = t_0 + \frac{i}{2A}\int |B|^2|\tilde{E}|^2 ( \bra{\rho_1}U^\dagger (\partial_\omega U)\ket{\rho_1} + \bra{\rho_2}U^\dagger (\partial_\omega U)\ket{\rho_2})d\omega.
     \label{eq:dt}
\end{equation}
The implicit assumption made here is that the pulse delay $\tau$ is very much shorter than the characteristic time scale for changes in the fibre birefringence, that is, $U$ is the same for both pulses. By direct substitution using the unitarity of $U$ and the orthogonality of $\vert \rho_i \rangle$ it can be shown that the integrand is zero, giving

%By direct substitution it can be easily shown that if $U$ is unitary and \langle \rho_i \vert \rho_j \rangle =\delta_{ij}
%since the rapid oscillation of $e^{i\omega \tau}$ under the integral gives zero for the cross terms. 
%and the integrand becomes zero, giving
\begin{equation}
     \langle t \rangle =  t_0.
     \label{eq:t0}
\end{equation}

\begin{figure}
    \centering
    \includegraphics[width=.47\linewidth]{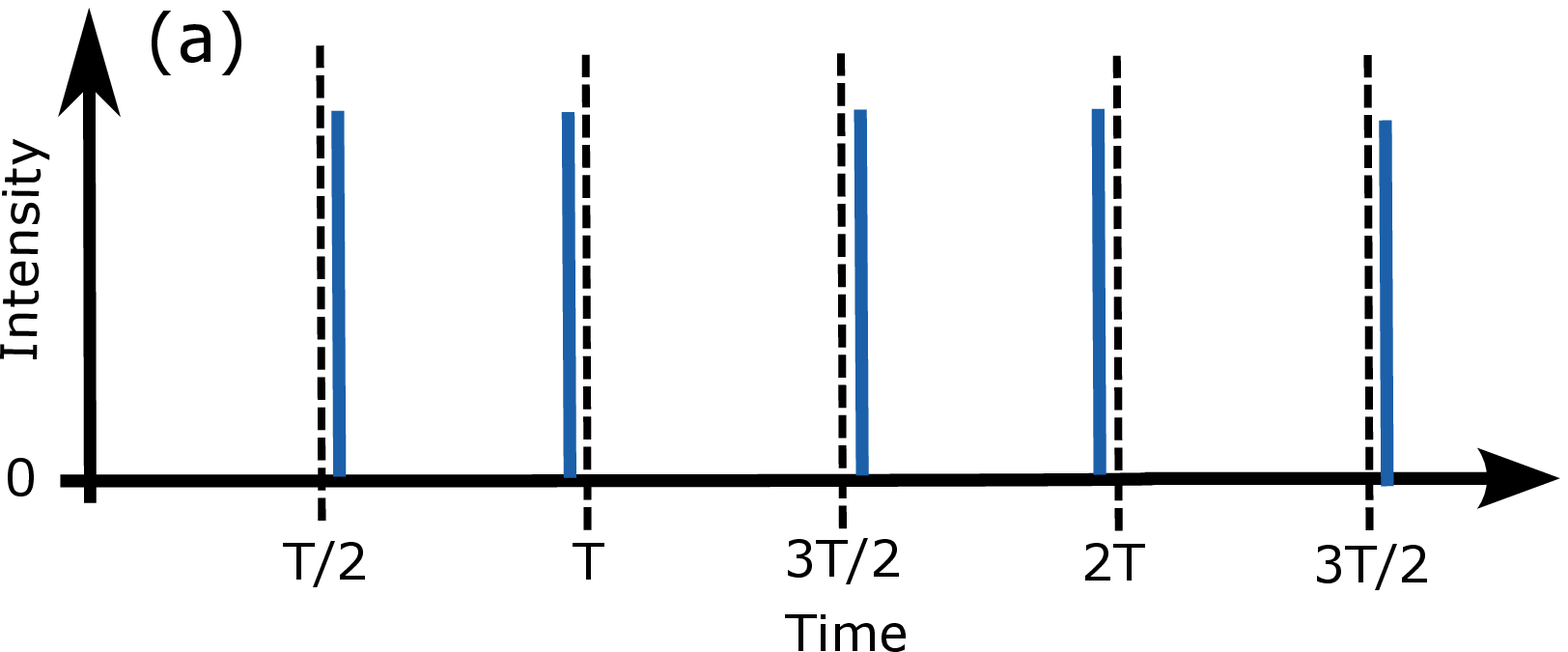}
    \hspace{.4cm}
    \includegraphics[width=.47\linewidth]{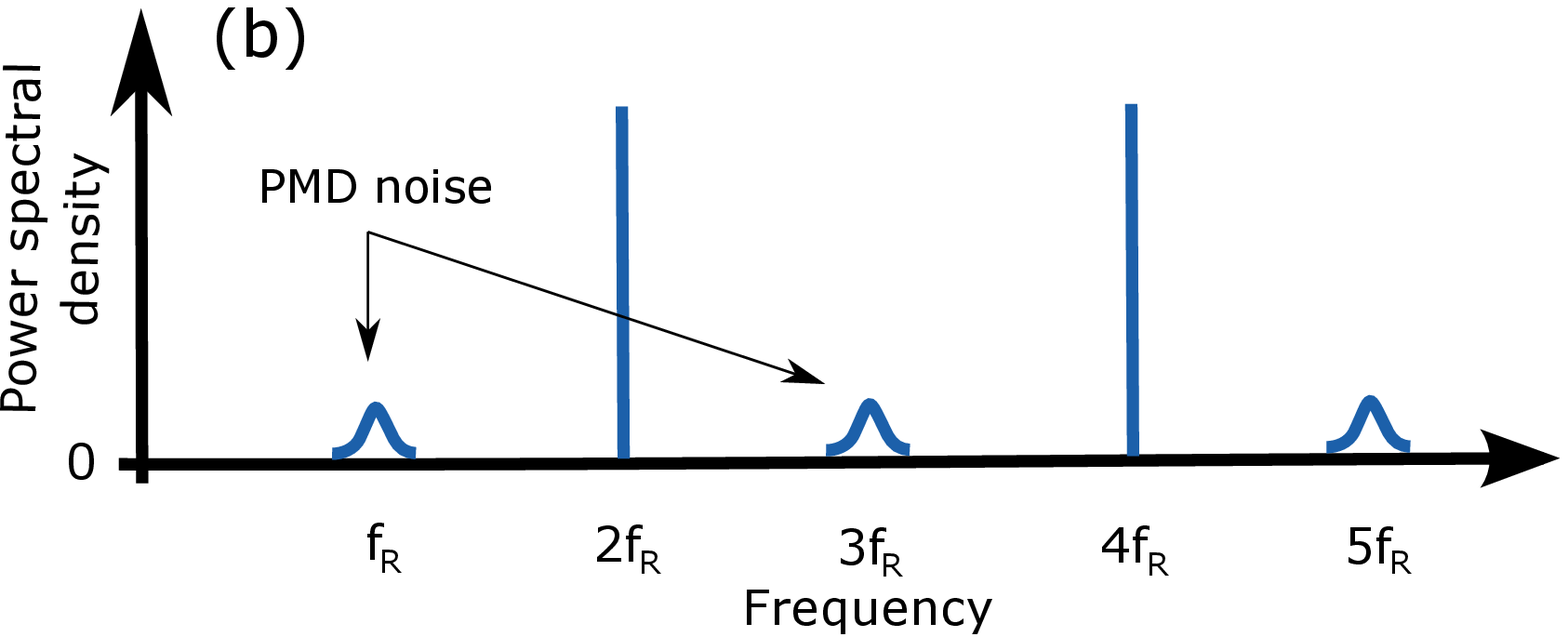}
    \caption{(a): Alternating between two orthogonal polarizations will alternatingly advance and retard the pulses with respect to the average. (b) In the RF spectrum, this timing noise will show up as sidebands between the fundamental spectral lines.}
    \label{fig:schematic}
\end{figure}

 This is an interesting result since it implies that for time and frequency distribution, polarization mode dispersion can be reduced without complicated control mechanisms simply by alternating between two orthogonal states of polarization and then using their average as the transfer signal. Fig. \ref{fig:schematic}a shows that in this case PMD will alternatingly advance and retard the pulses with respect to the average. The timing noise will thus appear at half the pulse repetition rate, creating sidebands in the RF spectrum between the fundamental lines. 
 
 In the next section a simple demonstration is presented where pulses from a femtosecond mode-locked laser with pulse repetition rate $f_R$ are time interleaved with orthogonal polarization and the receiver measures the phase at $2f_R$. %, reducing PMD by more than two orders of magnitude.
\begin{figure}
    \centering
    \includegraphics[width=.65\linewidth]{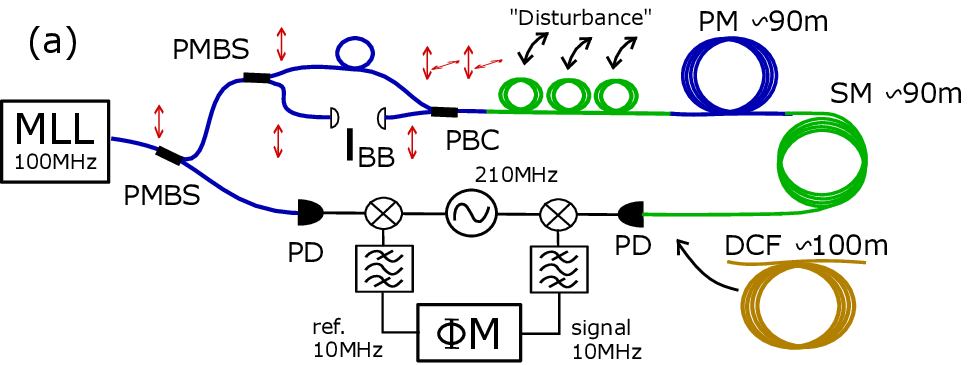}
    \includegraphics[width=.33\linewidth]{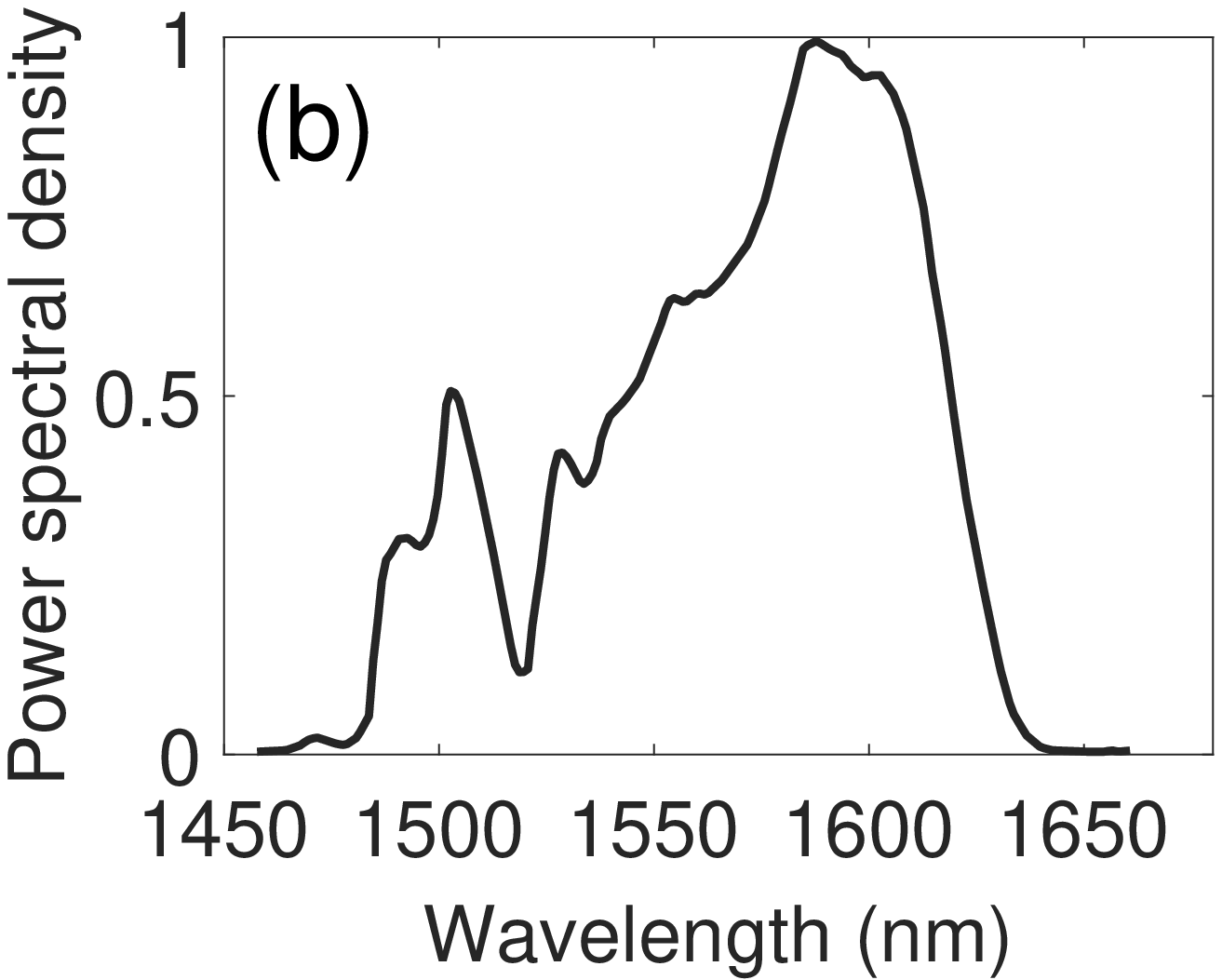}
    \caption{(a) Setup for testing PMD cancellation by alternating between two orthogonal, linear polarizations. MLL: mode-locked laser (100~MHz); PMBS: polarization-maintaining (PM) beam splitter; BB: beam block; PBC: polarizing beam combiner; DCF: dispersion compensating fibre; PD: photodetector; $\Phi$M: phase meter. PM fibre sections are blue, single-mode (SM) fibre sections are green. (b) Spectrum of the mode-locked laser.}
    \label{fig:setup}
\end{figure} 

\section{Experiment}

A schematic of the experimental setup is shown in Fig. \ref{fig:setup}. A free-running 100~MHz laser mode-locked by non-linear polarization rotation is used as light source, the broad spectrum of which is shown to the right (pulse transform limit $\sim70$~fs).
%The pulse spectrum is broad, and the transform-limited pulse duration is XX~fs. 
 
The pulse train is first polarized and then split in a 50/50 polarization maintaining (PM) beam splitter (PMBS). The first branch is sent to a reference photodetector (PD). The second branch is further split in two also in a 50/50 PM splitter. One of the branches is delayed with respect to the other and then the branches are recombined with orthogonal polarization in a polarizing beam combiner (PBC).
%The pulse train is first polarized and then split in two. The first branch is sent to a reference photodetector.
%The second branch is further split in two, one of the branches is delayed with respect to the other and then the branches are recombined with orthogonal polarizations in a polarizing beam combiner. 
 For experimental convenience, a free-space section is used for adjustment of the pulse delay and it also enables easy switching between single and dual polarization using a beam block. 

A manual 3-paddle fiber-optic polarization rotation stage consisting of three fiber loops  is used for varying the polarization (approx. delays $\frac{\lambda}{4},\frac{\lambda}{2},\frac{\lambda}{4}$). The pulses are then sent to a delay line consisting of a $\sim$90-m-long (installed) polarization maintaining (PM) fibre, the beat length of which is $\lesssim $5~mm, which means that the maximum polarization dependent delay change is nearly 100~ps. The signal is then returned via a $\sim$90-m-long single-mode (SM) fiber and optionally connected either to a 100-m dispersion compensating fibre spool (DCF) or directly to a photodetector. This type of test setup is short enough not to require active stabilization of the path length but still contains a lot of birefringence enabling the PMD effect to be easily investigated.

The pulse train signals at $f_\textrm{L}=200$~MHz are mixed with $f_{\textrm{LO}}=210$~MHz and the phase difference of the resulting 10~MHz signals are measured with a phase meter (Symmetricom 3120A). The measured time deviation $\Delta t_{M}$ is given by $\Delta t_{M} = f_{\textrm{L}}(f_{\textrm{LO}}-f_{\textrm{L}})^{-1}\Delta t$, which means that the real time deviation $\Delta t$ is magnified by a factor of 20. In the following, all results are presented in terms of the real time delay.

\section{Results}

\begin{figure}
    \centering
    \includegraphics[width=.49\linewidth]{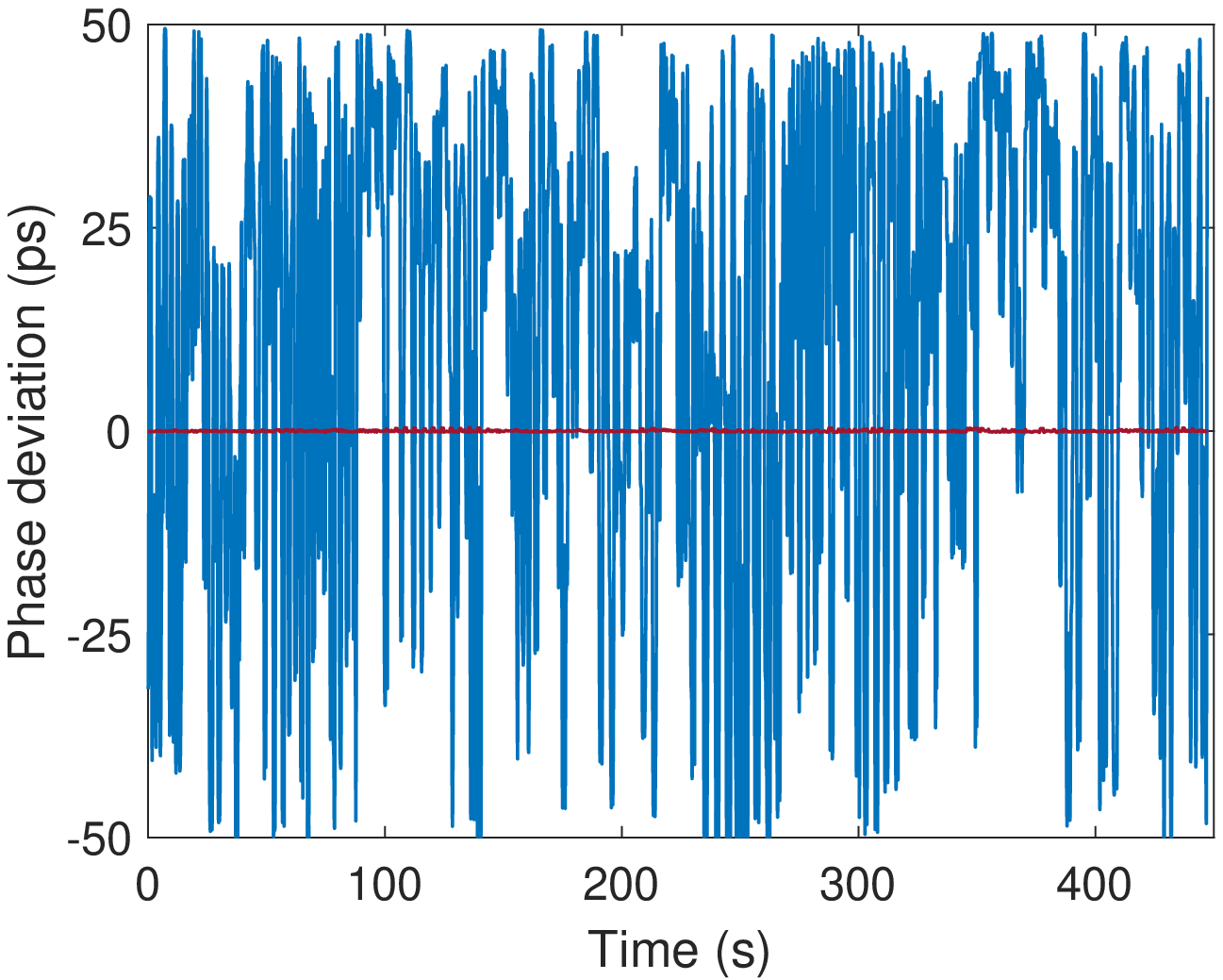}
      \includegraphics[width=.49\linewidth]{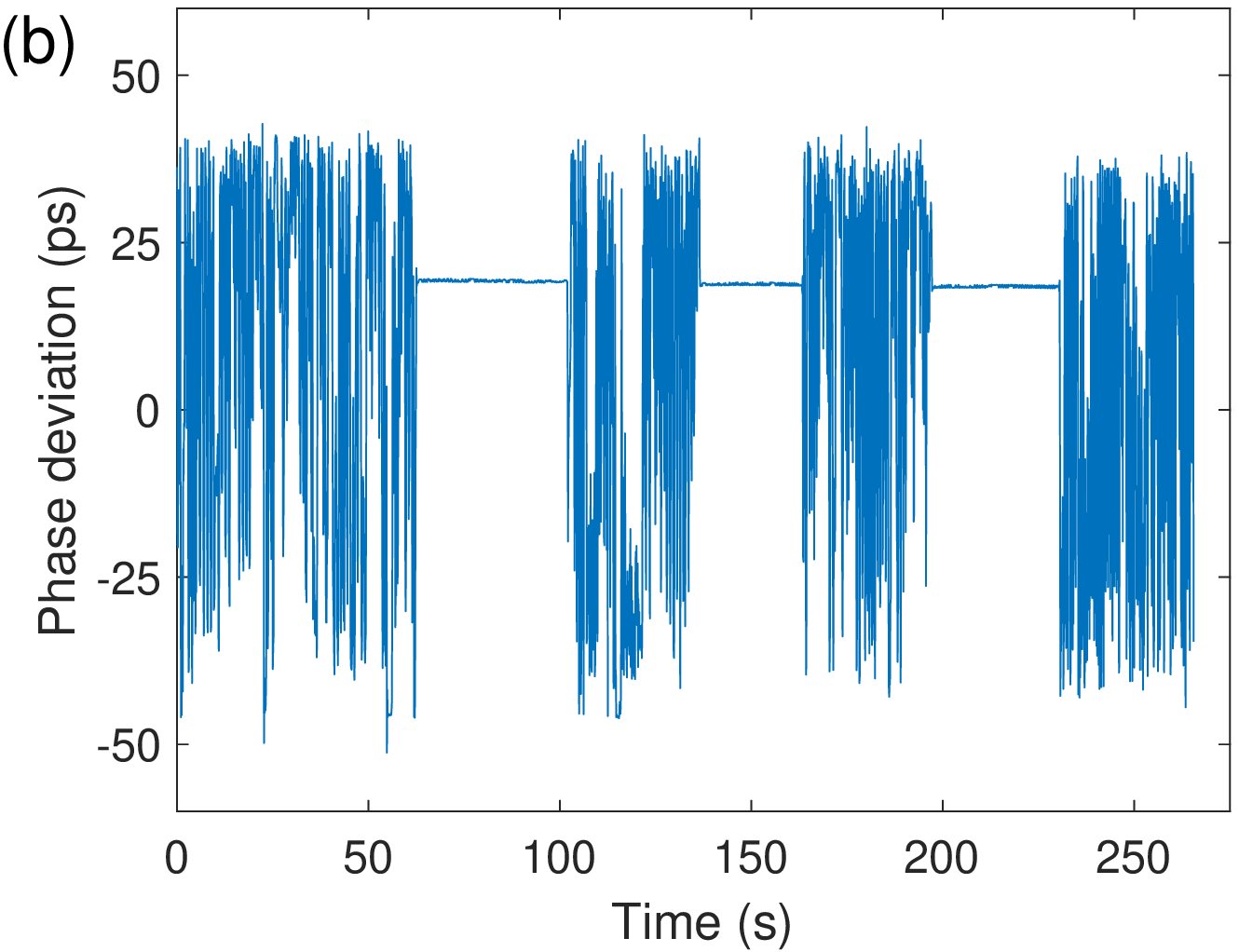}
    \caption{(a): Phase deviation with one  (blue) and two  alternating orthogonal polarizations (red) when the polarization rotators are manually rotated. (b) Blocking and unblocking one of the beams as a piece of SM fiber after the polarization rotator is manually disturbed.}
    \label{fig:timeseries}
\end{figure}

Delay variations for a single (blue) and alternating orthogonal (red) polarizations as the polarization rotators are manually turned (individually as well as simultaneously in a semi-random pattern) are show in Fig. \ref{fig:timeseries}a. Using orthogonal polarizations reduces the peak to peak delay variation from 103~ps to 600~fs. Fig. \ref{fig:timeseries}b shows similar data, but this time a piece of SM fiber after the beam combiner is disturbed as the beam block is intermittently put into the beam, again demonstrating the large reduction in PMD induced delay variations when two polarizations are used. 

\begin{figure}
    \centering
    \includegraphics[width=.49\linewidth]{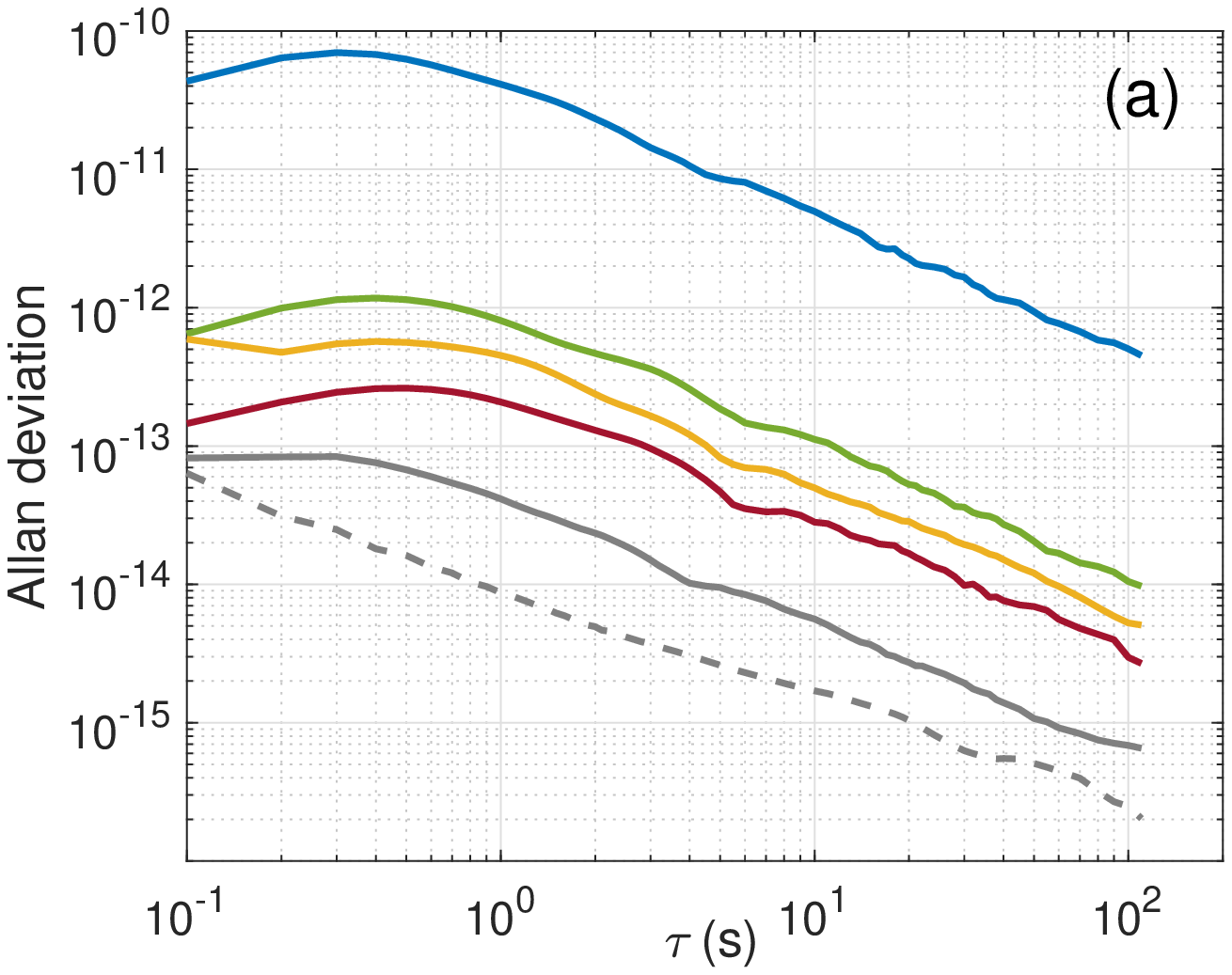}
    \includegraphics[width=.49\linewidth]{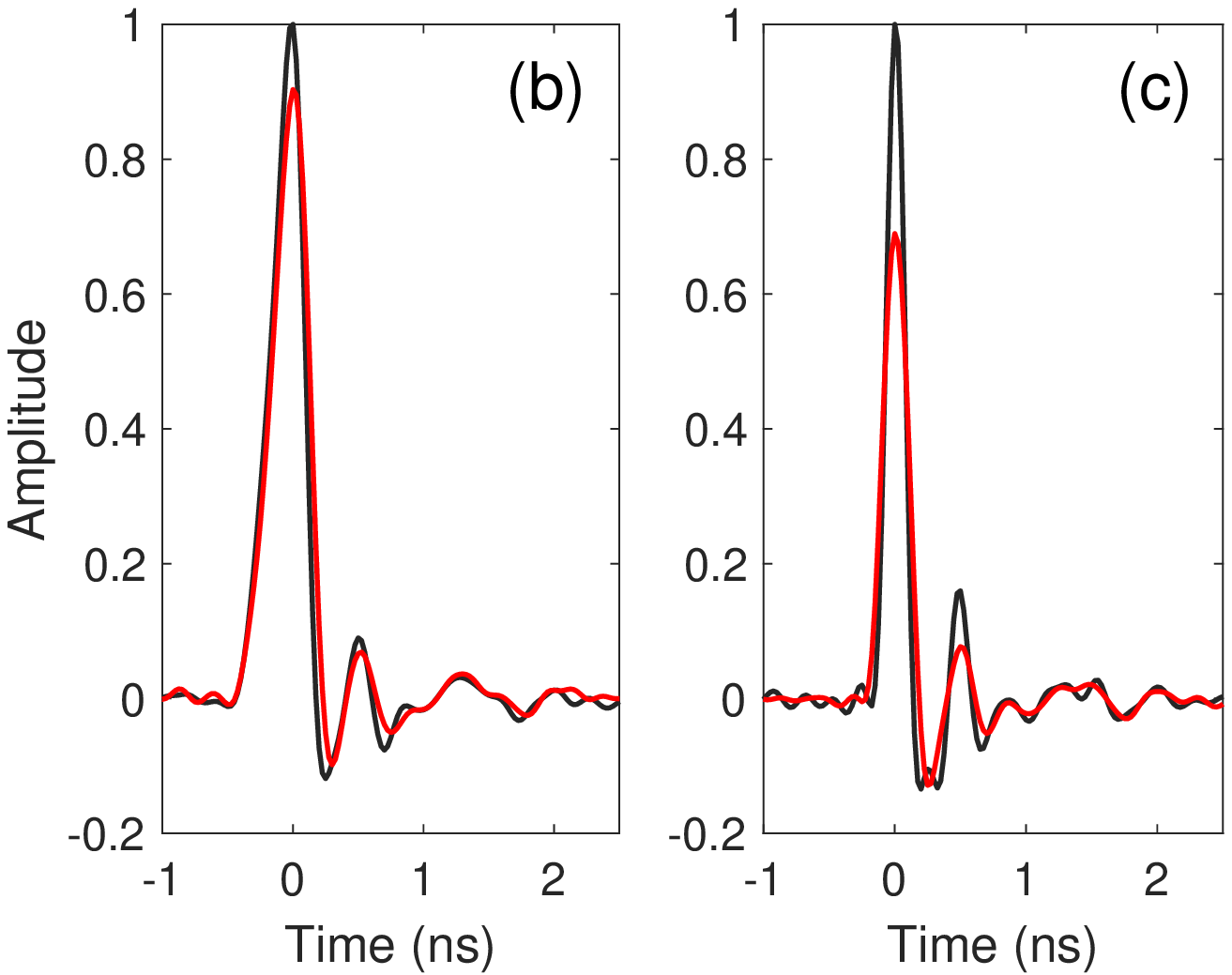}
      \includegraphics[width=.49\linewidth]{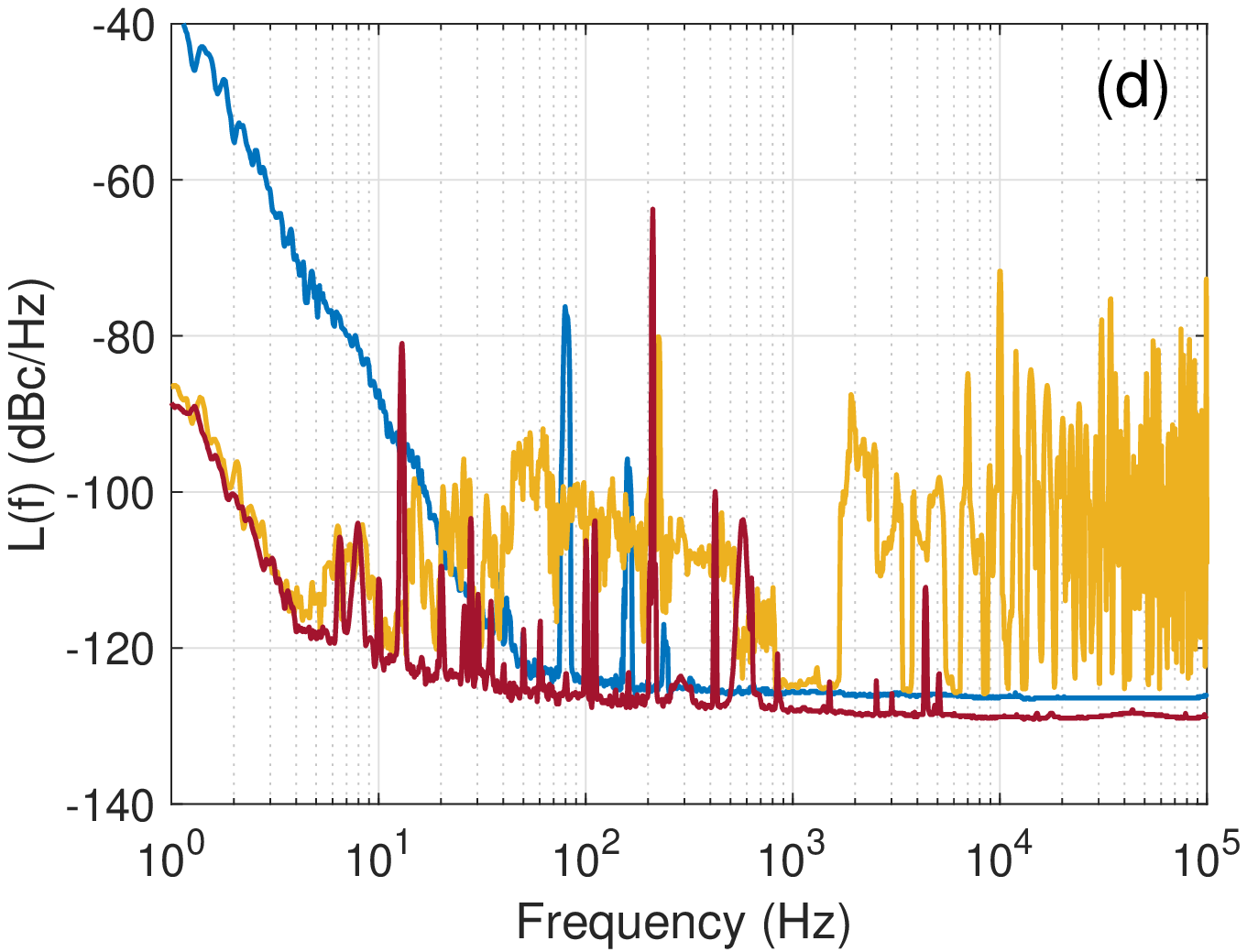}
    \caption{(a) Allan deviation with a single polarization (blue), alternating orthogonal with DCF (green), polarization scrambler without DCF (yellow) and alternating orthogonal without DCF (red). The gray curves show the adev with (solid) and without (dashed) perturbations when the fibre loops are bypassed. (b) Pulse amplitude variation (maximum and minimum) without DCF. (c) Pulse amplitude variation with DCF. (d) Single sideband phase noise with a single-polarization (blue), alternating orthogonal (red) and with polarization scrambling.}
    \label{fig:adev}
\end{figure}

Measured Allan deviations (adev) are show in Fig. \ref{fig:adev}a. The blue trace shows the adev with one polarization and the red with alternating orthogonal polarization without the DCF, showing a reduction by more than two orders of magnitude. When the DCF is used to compress the pulses (green), the PMD suppression becomes surprisingly somewhat worse. A possible reason for this can be found when comparing the pulse waveforms without (Fig. \ref{fig:adev}b) and with (Fig. \ref{fig:adev}c) the DCF. In the former case, the electrical pulse duration varies between 280~ps and 320~ps and the amplitude variation is about $\pm$5~\%. %, and is limited due to uncompensated dispersion and the broad spectrum of the source (Fig. \ref{fig:setup}). 
In the latter case the variation is 150-220~ps, which is limited from below by the oscilloscope and detector used to record the traces. The amplitude variation is $\pm$20~\%. This variation is expected and it is not a problem in itself; however, any non-linearities present in the measurement setup (photodiode, filters, mixers or amplifiers) may turn the large variations into phase fluctuations. Shorter pulses also mean that high-frequency reflections in the RF chain become more problematic, an issue which is further enhanced by the relatively larger pulse width changes. Measurements were also made without the DCF but with filtering the spectrum in order to reduce the pulse duration of the highly chirped pulses. In this case the amplitude variation was in between the two cases discussed above as was the PMD reduction. Several additional tests were made, including testing different photodetectors (5~GHz and 45~GHz) and varying the photodiode bias voltage, but at present the root cause remains unclear, unfortunately. Finally, in Fig. \ref{fig:adev}a, the yellow trace shows the adev when a polarization scrambler based on resonant piezo-based fibre squeezers operating at 700~kHz is used (General Photonics PCD-104, \cite{Yao2002a}).
%Therefore, keeping the PMD small compared to the optical pulse duration will keep the amplitude variations small. Also, with large changes in the pulse duration, there will be significant changes amplitudes of the harmonics will change 
%Measurements were also made without the DCF but with filtering the spectrum in order to reduce the pulse duration. In this case the amplitude variation was in between the two  examples shown as was the PMD compensation.  
%as recorded on a. With the DCF, the pulse amplitude variations are much stronger due to the   
%Possible reasons for this are discussed in the next section. 

Figure \ref{fig:adev}d shows the recorded single-sideband phase noise spectra. With two polarizations (red) the phase noise at low frequencies, which  correspond to the applied perturbations, is significantly reduced compared to the single polarization (blue). The scrambler shows a similar reduction at low frequencies, but in stark contrast to the dual polarization scheme it also induces a considerable amount of additional phase noise starting very close to the carrier (10~Hz) even though the fiber squeezers operate at a much higher rate (700~kHz) .

\section{Discussion}

The results show that using alternating orthogonal polarizations can be very effective in combating PMD. Compared to using scramblers, three advantages can be seen: (i) PMD noise appears intrinsically far away from the carrier; (ii) no additional modulators are needed when using (non-overlapping) pulses as in this work; (iii) the polarization at the transmitter is well defined, which means that when the light is reflected back from a distant receiver using a Faraday mirror, high-contrast interference will take place just as with single polarization. This means that optical phase detection, e.g.,  balanced-optical cross-correlation \cite{Kim2007a, Xin2018a, Safak2018a} or spectral interferometry \cite{Marra2012a}, can be used for phase detection for active phase compensation of the fiber. Note that in this case, the feedback must act according to the average phase shift between the two polarizations. With high-speed scramblers, the scrambler will have time to change state during the round-trip time, making the polarization state less clear at the transmitter unless the scrambling frequencies are chosen carefully. %Future work should include investigations of the use of amplitude insensitive RF phase detection as well as optical phase detection in closed loop. Also replacing the MLL with low-cost, narrowband, telecom SFP transmitters as well as long-haul transmission should be very interesting.

The test setup used a mode-locked laser, but standard telecom SFP transceivers should also work as long as the off-state has sufficiently high attenuation not to affect appreciably the orthogonality of the interleaved pulses. Custom made transceivers with, e.g., two built-in electro-absorption modulators for two orthogonal polarizations or an additional phase modulator for polarization switching would remove the need for an external interleaver \cite{Xie2004a}.

What is troublesome about PMD for accurate, long-term time transfer, is that even though there is considerable PMD present, it will not show up in the transfer stability as long as the fibers remain still. Even a long test run might not show the actual vulnerability to PMD unless the link is perturbed properly during testing. For instance, in the test setup discussed above, even with a single polarization and without securing the fibers anywhere, the link can operate for hours without any indication of instability issues, but the very slightest touch to the fiber can induce a phase shift of almost 100~ps.

If the fibers are sufficiently still as might be the case with buried fibers, the PMD will mostly affect the very long-term stability and thus might be less of an issue for frequency transfer. To reach, e.g., remote scientific experiments, fibers might need to be hanging in telephone poles or resting on the ground, in which case wind gusts and severe temperature changes can induce short-term instability.  For example, on a 2x50~km fibre link in \cite{Fordell2020a}, in addition to a slow drift of the polarization on a few minute time scale, intermittent fast oscillations with a period of 4~Hz were observed that caused oscillations in the adev at the $10^{-13}$ level. This instability was most likely caused by oscillations of an airborne piece of the fiber link. In situations like this, the use of alternating orthogonal polarization should be a simple and effective way of mitigating the PMD problem without inducing additional phase noise close to the carrier.

%\begin{figure}
  %  \centering
  %  \includegraphics[width=.49\linewidth]{pulsesCombined.eps}
  %    \includegraphics[width=.49\linewidth]{spectrum.eps}
  %  \caption{(a): Phase deviation with one  (blue) and two alternating %orthogonal polarizations when the polarization rotators are manually %rotated. (b) Blocking and unblocking one of the beams as a piece of SM %fiber after the polarization rotators is manually disturbed. The phase %excursion is smaller than using the polarization rotators.}
%    \label{fig:pulses}
%\end{figure}

%Especially when the fibres are not buried but are airborne, e.g., hanging in telephone poles, the PMD issue can be significant \cite{Fordell2020a}. Polarization mode scrambles can be used to mitigate the problem \cite{FrenchLink}\, but high-speed scramblers (<1~MHz) are quite expensive and they still produce noise close to the carrier.

%CLEAN UP optical OSCILLATOR - how?
%orthogonal input=> orthogonal output

\section{Conclusions}
In conclusion, the use of optical pulses with alternating orthogonal polarization was introduced as a potentially simple and effective way of mitigating polarization mode dispersion in fibre-optic time and frequency transfer. When pulses are used, no additional modulators are needed; a simple passive pulse interleaver is sufficient. The test setup used a mode-locked laser but continuous-wave light sources using amplitude modulators, such as standard telecom SFP transceivers, should work as well. Custom transceivers would remove the need for an external interleaver. The test setup demonstrated a significant reduction in PMD in a short fibre link with high PMD, but more work on the performance limits is needed. Especially, the influence of polarization-dependent loss (or gain) should be investigated as well as operation of the setup over long-haul links and in closed loop so that non-PMD related phase drifts would be compensated simultaneously.

\begin{backmatter}
\bmsection{Funding}
Academy of Finland (grants 296476 and 328389).

\bmsection{Acknowledgments}
The work is part of the Academy of Finland Flagship Programme, Photonics Research and Innovation (PREIN), decision 320168. 

\bmsection{Disclosures}
The authors declare no conflicts of interest.

\bmsection{Data availability} 
Data underlying the results presented in this paper are not publicly available at this time but may be obtained from the author upon reasonable request.

%\bmsection{Supplemental document}
%See Supplement 1 for supporting content. 

\end{backmatter}

%%%%%%%%%%%%%%%%%%%%%%% References %%%%%%%%%%%%%%%%%%%%%%%%%

%Add references with BibTeX or manually.

%%%%%%%%%% If using BibTeX:
\bibliography{sample}

%%%%%%%%%% If preparing manually:
% \newcommand{\enquote}[1]{``#1''}

% \bibitem{Zhang:14}
% Y.~Zhang, S.~Qiao, L.~Sun, Q.~W. Shi, W.~Huang, L.~Li, and Z.~Yang,
%   \enquote{Photoinduced active terahertz metamaterials with nanostructured
%   vanadium dioxide film deposited by sol-gel method,}
%   {\protect\JournalTitle{Optics Express}} \textbf{22}, 11070--11078 (2014).

% {Optical Society}, \enquote{{OSA Publishing},}
%   \url{http://www.osapublishing.org}.

% \bibitem{FORSTER2007}
% P.~Forster, V.~Ramaswamy, P.~Artaxo, T.~Bernsten, R.~Betts, D.~Fahey,
%   J.~Haywood, J.~Lean, D.~Lowe, G.~Myhre, J.~Nganga, R.~Prinn, G.~Raga,
%   M.~Schulz, and R.~V. Dorland, \enquote{Changes in atmospheric consituents and
%   in radiative forcing,} in \enquote{Climate Change 2007: The Physical Science
%   Basis. Contribution of Working Group 1 to the Fourth assesment report of
%   Intergovernmental Panel on Climate Change,}  S.~Solomon, D.~Qin, M.~Manning,
%   Z.~Chen, M.~Marquis, K.~B. Averyt, M.~Tignor, and H.~L. Miler, eds.
%   (Cambridge University Press, 2007).

% \end{thebibliography}

\end{document}